 \documentclass[12pt]{article} 
 \usepackage{color} 
 \newcommand{\comma}{\;\; ,}
 \newcommand{\period}{\;\; .}
 \newcommand{\eq}{\; = \;}
 \newcommand{\sep}{\;\; , \;\;}
 \newcommand{\be}{\begin{equation}}
 \newcommand{\bd}{\begin{displaymath}}
 \newcommand{\ee}{\end{equation}}
 \newcommand{\ed}{\end{displaymath}}
 \newcommand{\ba}{\begin{eqnarray}}
 \newcommand{\ea}{\end{eqnarray}}
 
 \newcommand{\minus}{\! - \!}
 \newcommand{\mod}{{\rm mod}\,}
 \newcommand{\plus}{\! + \!}
 \newcommand{\del}{\!\!\!\!\!}
 \renewcommand{\i}{{\rm i}}
 \newcommand{\e}{{\rm e}}

 \title{ Hyperelliptic parametrization of the 
 generalized order parameter of the $N=3$ chiral Potts model }

 \author{ R.J. Baxter\\
 {\protect \small  Mathematical
 Sciences Institute}\\
 {\protect  \small The Australian National University,
 Canberra, A.C.T. 0200,
  Australia}}

 \date{\protect \small  6 January 2006}

\begin{document}


 \maketitle

 \abstract{ It has been known for some time that the
 Boltzmann weights of the chiral Potts model can be parametrized
 in terms of hyperelliptic functions, but as yet no such 
 parametrization has been applied to the partition and correlation
 functions. Here we show that for $N=3$ the function $S(t_p)$
 that occurs in the recent calculation of the order parameters
 can be expressed quite simply in terms of such a parametrization. }






 \section{Introduction}

 There are a few two-dimensional models (and even fewer 
 three-dimensional models) in equilibrium statistical 
 mechanics that have been solved exactly. These are lattice models
 where spins $\sigma_i$ are assigned to the sites $i$ of a lattice 
 (usually the square lattice). Each spin takes one of $N$ possible 
 values
 and spins  $\sigma_i,  \sigma_j$ on adjacent sites $i,j$ interact 
 with a specified positive real Boltzmann weight function 
 $W(\sigma_j,\sigma_j)$.
 One wants to calculate the partition function (also called the 
 sum-over-states)
 \be Z \eq \sum \prod_{<ij>} W(\sigma_j,\sigma_j) \comma \ee
 where the sum is over all states of all the spins, and the product
 is over all edges $(i,j)$ of the lattice.

 If the number of sites is $M$, we expect the limit
 \be \kappa = \lim_{M \rightarrow \infty} Z^{1/M} \ee
 to exist and to independent of the shape of the lattice,
 provided it becomes large in all directions: this is the 
 ``thermodynamic limit'', and $\kappa$ is the exponential of the 
 free energy per site.  If $1, \ldots , m$
 are  sites fixed on the lattice and the limit is taken so they
 become infinitely deep within it, then we also expect the average
 \be \langle f(\sigma_1, \ldots, \sigma_m ) \rangle \eq Z^{-1} 
 \sum \prod_{<ij>} f(\sigma_1, \ldots, \sigma_m ) 
 W(\sigma_j,\sigma_j)  \ee
 to tend to a limit, for any given function $f$ of these $m$ spins.

 Because spins only interact with their neighbours, 
 one can build up the lattice one row at a time, and associate a 
 row-to-row ``transfer matrix'' with such an operation.

 To solve such a model, typically one shows that the Boltzmann 
 weights $W$ satisfy the star-triangle or ``Yang-Baxter''  
 relations \cite{book82}. 
 These ensure certain commutation relations between the  transfer
 matrices, and this is usually a first step towards calculating 
 $\kappa$.

 The next step is to calculate the order parameters, which are 
 averages of certain functions of a single spin $\sigma_1$ deep 
 within the lattice. This is a harder problem than calculating 
 $\kappa$. For instance, Onsager \cite{Ons44} calculated $\kappa$ 
 for the square-lattice Ising model in 1944 , but it was not till 
 1949  that he
 announced at a conference his result for the order parameter
 (namely the spontaneous magnetization), and not till 1952 
 before a proof of the result  was published by Yang \cite{Yang52} .

 However, since then the ``corner transfer matrix'' method
 has been developed by Baxter \cite{RJB81}, and the ``broken rapidity 
 line
 method'' by Jimbo, Miwa and Nakayashiki \cite{JMN93}. For many of the 
 solved models (those with the ``rapidity difference'' property),
 these methods make the calculation of the order parameters
 comparitively straighforward.

 Even so, one model has proved challenging, namely
 the chiral Potts model. This is an $N$-state model where
 $W(\sigma_j,\sigma_j)$ depends only on the spin difference 
 $\sigma_j - \sigma_j$, mod $N$. The Boltzmann weights
 also depend on two parameters $p, q$, (known as 
 ``rapidities''), and on given positive real constants $k, k'$, 
 related by
 \be \label{kk'}
 k^2 + {k'}^2 = 1 \period \ee
 The parameter $k'$ plays the role of a temperature, being
 small at low temperatures. For $ 0 < k' <1 $ the system 
 displays spontaneous ferromagnetic order, becoming critical
 as $k' \rightarrow  1$.

 Its order parameters can be taken
 to be
 \be \label{Mr}
 {\cal M}_r \eq \langle \omega^{r \sigma_1} \rangle \comma \ee
 where $\omega = \exp( 2 \pi \i/N)$ and $r = 1, \ldots , N-1$. It 
 was shown in 1988 that its Boltzmann weights satisfy the 
 star-triangle relation \cite{BPAuY88,AuYP89}, and the partition 
 function per site $\kappa$ was soon calculated \cite{RJB88,RJB90}.

 The order parameters were another story. The model had developed 
 from a one-dimensional quantum spin chain, which has the same order 
 parameters. From series expansions it was conjectured 
 \cite{AMPT89} in 1989  that 
 \be \label{conj}
 {\cal M}_r \eq k^{r(N-r)/N^2} 
 \period \ee
 Much effort was expended in the ensuing years (certainly by the  
 author) in attempting to derive this result. It was not until
 2005 that this was done \cite{RJB05a,RJB05b}.


 \setlength{\unitlength}{1pt}
 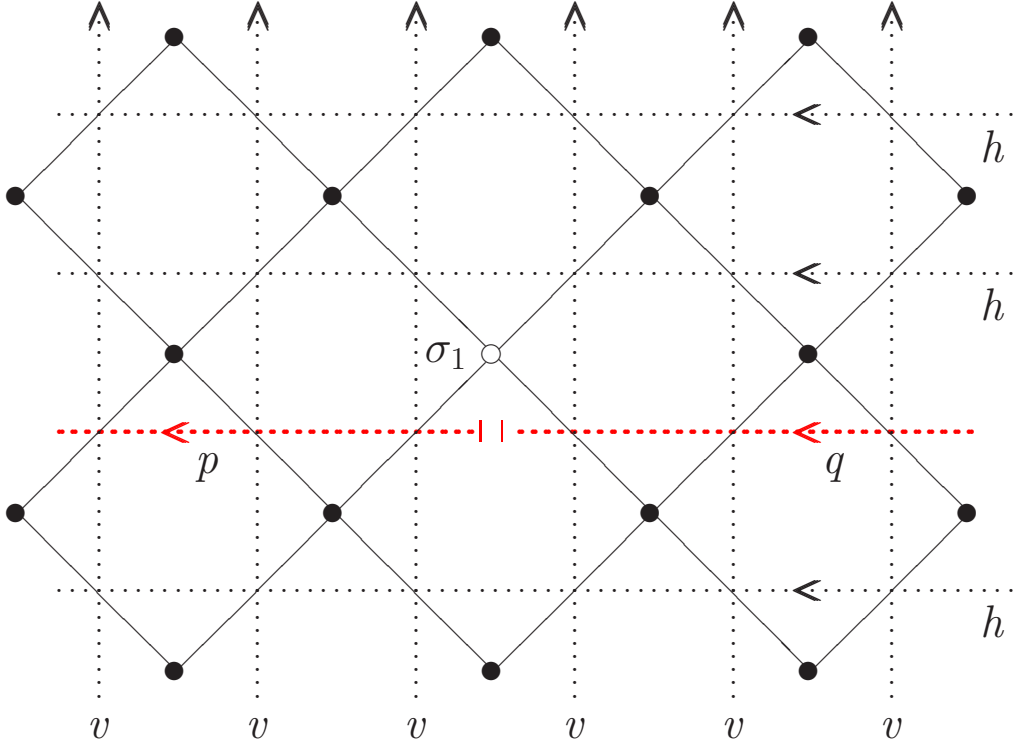
\begin{figure}[hbt]

 \begin{picture}(420,260) (0,0)

 \multiput(30,15)(5,0){73}{.}
 \multiput(26,75)(5,0){32}{\color{red}{ \bf .}}
 \multiput(27,75)(5,0){32}{\color{red}{ \bf .}}
 \multiput(200,75)(5,0){35}{\color{red}{ \bf .}}
 \multiput(201,75)(5,0){35}{\color{red}{ \bf .}}
 \multiput(30,135)(5,0){73}{.}
 \multiput(30,195)(5,0){73}{.}
 \put (191,72) {\color{red}{\line(0,1) {8}}}
 \put (199,72) {\color{red}{\line(0,1) {8}}}
 \thicklines

 \put (69,72) {\color{red}{\large $< $}}
 \put (70,72) {\color{red}{\large $< $}}
 \put (71,72) {\color{red}{\large $< $}}

 \put (308,12) {\large $< $}
 \put (309,12) {\large $< $}
  \put (310,12) {\large $< $}

 \put (308,72) {\color{red}{\large $< $}}
 \put (309,72) {\color{red}{\large $< $}}
 \put (310,72) {\color{red}{\large $< $}}

 \put (308,132) {\large $< $}
 \put (309,132) {\large $< $}
 \put (310,132) {\large $< $}

 \put (308,192) {\large $< $}
 \put (309,192) {\large $< $}
 \put (310,192) {\large $< $}

 \put (42,230) {\large $\wedge$}
 \put (42,229) {\large $\wedge$}
 \put (42,228) {\large $\wedge$}

 \put (102,230) {\large $\wedge$}
 \put (102,229) {\large $\wedge$}
 \put (102,228) {\large $\wedge$}

 \put (162,230) {\large $\wedge$}
 \put (162,229) {\large $\wedge$}
 \put (162,228) {\large $\wedge$}

 \put (222,230) {\large $\wedge$}
 \put (222,229) {\large $\wedge$}
 \put (222,228) {\large $\wedge$}

 \put (282,230) {\large $\wedge$}
 \put (282,229) {\large $\wedge$}
 \put (282,228) {\large $\wedge$}

 \put (342,230) {\large $\wedge$}
 \put (342,229) {\large $\wedge$}
 \put (342,228) {\large $\wedge$}

 \thinlines


 \put (170,102) {{\Large $\sigma_1$}}
 \put (320,60) {{\Large \it q}}
 \put (83,60) {{\Large \it p}}
 \put (380,-2) {{\Large \it h}}
 \put (380,118) {{\Large \it h}}
 \put (380,178) {{\Large \it h}}

 \put (195,105) {\circle{7}}

 \put (16,45) {\line(1,-1) {60}}
 \put (16,165) {\line(1,-1) {180}}
 \put (76,225) {\line(1,-1) {117}}
 \put (198,103) {\line(1,-1) {117}}
 \put (196,225) {\line(1,-1) {180}}
 \put (316,225) {\line(1,-1) {60}}
 \put (16,165) {\line(1,1) {60}}
 \put (16,45) {\line(1,1) {180}}
 \put (76,-15) {\line(1,1) {117}}
 \put (198,107) {\line(1,1) {118}}
 \put (196,-15) {\line(1,1) {180}}
 \put (316,-15) {\line(1,1) {60}}

 \put (75,105) {\circle*{7}}
 \put (315,105) {\circle*{7}}
 \put (75,-15) {\circle*{7}}
 \put (195,-15) {\circle*{7}}
 \put (315,-15) {\circle*{7}}

 \put (15,45) {\circle*{7}}
 \put (135,45) {\circle*{7}}
 \put (255,45) {\circle*{7}}
 \put (375,45) {\circle*{7}}

 \put (15,165) {\circle*{7}}
 \put (135,165) {\circle*{7}}
 \put (255,165) {\circle*{7}}
 \put (375,165) {\circle*{7}}

 \put (75,225) {\circle*{7}}
 \put (195,225) {\circle*{7}}
 \put (315,225) {\circle*{7}}

 \put (42,-40) {{\Large \it v}}
 \put (102,-40) {{\Large \it v}}
 \put (162,-40) {{\Large \it v}}
 \put (222,-40) {{\Large \it v}}
 \put (282,-40) {{\Large \it v}}
 \put (342,-40) {{\Large \it v}}

 \multiput(45,-25)(0,5){52}{.}
 \multiput(105,-25)(0,5){52}{.}
 \multiput(165,-25)(0,5){52}{.}
 \multiput(225,-25)(0,5){52}{.}
 \multiput(285,-25)(0,5){52}{.}
 \multiput(345,-25)(0,5){52}{.}
 \end{picture}

 \vspace{1.5cm}
 \caption{\footnotesize  The square lattice (circles and solid lines, 
 drawn diagonally) and its medial graph of dotted or broken lines.}
 \label{Fpq}
 \end{figure}

 The method used was based on that of Jimbo {\it et al} \cite{JMN93}.
 In Figure \ref{Fpq} we show the square lattice $\cal L$, drawn 
 diagonally, denoting the sites by circles and the edges by solid lines. 
 We also show as dotted (or broken) lines the medial graph of  $\cal L$. 
 Every edge of  $\cal L$ is intersected by two dotted lines. With each 
 dotted line we associate a rapidity variable ($p$, $q$, $h$ or $v$). 
 In general these variables  may differ from dotted line to dotted line. 
 They must be the same all along the line, except for the horizontal 
 broken line immediately below the central  spin $\sigma_1$. We break 
 this below $\sigma_1$ and assign a rapidity $p$ to the left of the 
 break, a rapidity $q$ to the right. With these choices of rapidities,
 define
 \be \tilde{F}_{pq} (r) \eq \langle \omega^{r \sigma_1} \rangle 
 \period \ee 


 In the thermodynamic limit, the star-triangle relations will ensure 
 that $\tilde{F}_{pq} (r) $ is independent of the `` background'' 
 rapidities
 $v, h$, because it allows us to move any of these dotted lines
 off to infinity.\cite{RJB78}  However, the effect of the break is 
 that we  cannot move the broken line $p,q$ away from $\sigma_1$, so
 $\tilde{F}_{pq} (r) $  will indeed depend on $p$ and $q$. 

 An important special case is when $q = p$. Then the $p,q$ rapidity line
 is not in fact broken, so it can be removed to infinity and
 $\tilde{F}_{pp} (r) $ must be independent of $p$ and equal to the 
 order parameter ${\cal M}_r$ defined by (\ref{Mr}):
 \be {\cal M}_r \eq \tilde{F}_{pp} (r) \period \ee

 We also define
 \be \label{defG}
 G_{pq}(r) \eq {\tilde F}_{pq}(r)/{\tilde F}_{pq}(r-1) \period \ee

 The author wrote down \cite{RJB98} functional relations satisfied by
 $G_{pq}(r)$ in 1998. They do not completely specify $G_{pq}(r)$,
 but must be supplemented by information on the analyticity 
 properties of $G_{pq}(r)$. (Just as the relation $f(z+1) = f(z)$
 only tells us that $f(z)$ is periodic of period 1: however, if we 
 can also show that $f(z)$ is analytic and bounded in the domain 
 $0 \leq \Re (z) < 1$, then it follows from Liouville's theorem
 that $f(z)$ is a constant.)

 For $N=2$ the chiral Potts model reduces to the Ising model and 
 it is quite easy to find the needed analyticity information, to 
 solve the functional relations and obtain the Onsager-Yang result
 ${\cal M}_1 = k^{1/4}$.

 For $N > 2$ the problem is much harder. It was not until late 2004 
 that the author realised that it is not actually necessary to solve
 for the general function  $G_{pq}(r)$. It is sufficient to do
 so for a special ``superintegrable'' case where $q$ is related to 
 $p$. The function then has quite simple analyticity properties 
 and it quite easy to solve the relations (in fact one does not 
 even need all the relations), to obtain $G_{pq}(r)$ for this case
 and to verify the 16--year old conjecture (\ref{conj}). For $r = 1,
 \ldots , N$, the functions $G_{pq}(r)$ can all be expressed
 in terms of a single function $S(t_p)$ which is defined below.

 Even so, it would still be interesting to understand  $G_{pq}(r)$
 more generally. A fundamental difficulty is that for $N > 2$ 
 the rapidities $p$ and $q$ are 
 points on an algebraic curve of genus greater than 2, and there is
 no explicit parametrization of this curve in terms of 
 single-valued functions of a single variable. (There is for 
 $N = 2$: one can then parametrize in terms of Jacobi elliptic 
 functions.) One can parametrize in terms of hyperelliptic functions
 \cite{RJB91}, but these have $N-1$ arguments that are related to one 
 another. As yet they have not proved particularly useful, but one 
 lives in hope. The function $S(t_p)$ is a simple example of a 
 thermodynamic property of the chiral Potts model, and has the 
 simplifying feature that it depends on only one rapidity, rather 
 than two. It is an interesting question whether it can be 
 simply expressed in terms of these hyperelliptic functions.

 For $N=3$ these hyperelliptic functions can be expressed in terms of
 ordinary Jacobi elliptic functions. One still has two related arguments
 (here termed $z_p$ and $w_p$), but some of the properties can be expressed
 as products of Jacobi functions, each with an argument $z_p$ or $w_p$, or 
 some combination thereof. A number of such results have been 
 obtained.\cite{RJB93a}, \cite[pp. 568, 569]{RJB93b}

 There are two distinct ways of performing the hyperelliptic 
 parametrization. In \cite{RJB91,RJB05c} we used what
 we shall herein call the ``original'' parametrization. What we report 
 here is that for $N = 3$ the function  $S(t_p)$ can be expressed 
 quite simply as a product of Jacobi functions of $z_p$ and $w_p$, 
 {\em provided}  we use the second ``alternative'' parametrization.






 \section{The function $S(t_p)$}

 We can take a rapidity  $p$ to be a set of variables
 $p = \{ x_p,y_p,\mu_p, t_p \}$ related to one another by
 \ba \label{prlns}   t_p = x_p y_p \sep & & \del \del 
 x_p^N + y_p^N = k(1+x_p^N y_p^N) 
 \comma \nonumber \\
 && \\
 k x_p^N = 1-k'/\mu_p^N 
 & , & k y_p^N = 1-k'\mu_p^N \period \nonumber \ea

 There are various automorphisms or maps that take one set 
 $\{ x_p,y_p,\mu_p, t_p \}$ to another set satisfying the same 
 relations (\ref{prlns}). Four that we shall use are:
 \ba \label{autos}
 R:  \{ x_{Rp},y_{Rp},\mu_{Rp},t_{Rp} \} & = &
 \{ y_p,\omega x_p,1/\mu_p, \omega t_p \} \comma \nonumber \\
 S:  \{ x_{Sp},y_{Sp},\mu_{Sp},t_{Sp} \} & = & 
 \{ y_p^{-1},x_p^{-1},\omega^{-1/2} y_p/(x_p \mu_p),  
 t_p^{-1} \} \comma \nonumber \\
 V:  \{ x_{Vp},y_{Vp},\mu_{Vp},t_{Vp} \} & = & 
 \{ x_p,\omega y_p, \mu_p, \omega t_p \} \comma  \\
 M:  \{ x_{Mp},y_{Mp},\mu_{Mp},t_{Mp} \} & = & 
 \{ x_p, y_p, \omega \mu_p, \omega t_p \} \period \nonumber \ea
 They satisfy
 \bd 
 R V^{-1} R = V  \sep M R M = R \sep M S M = S  \comma \ed
 \be \label{Veqns} S^2 =  V^N = M^N =  1   \period \ee

 Let $q$ be another rapidity set, related to $p$ by
 $q = Vp$, i.e. 
 \be \label{qp}
 x_q = x_p \sep y_q = \omega y_p \sep \mu_q = \mu_p \period \ee

 We take $\mu_p$ to be outside the unit circle, so
 \be \label{mupr}
 |\mu_p | > 1 \period \ee 
 Then we can specify
 $x_p$ uniquely by requiring that
 \be \label{xpr}
 -\pi/(2 N) < \arg (x_p ) < \pi/(2 N) \period \ee


 \setlength{\unitlength}{1pt}
 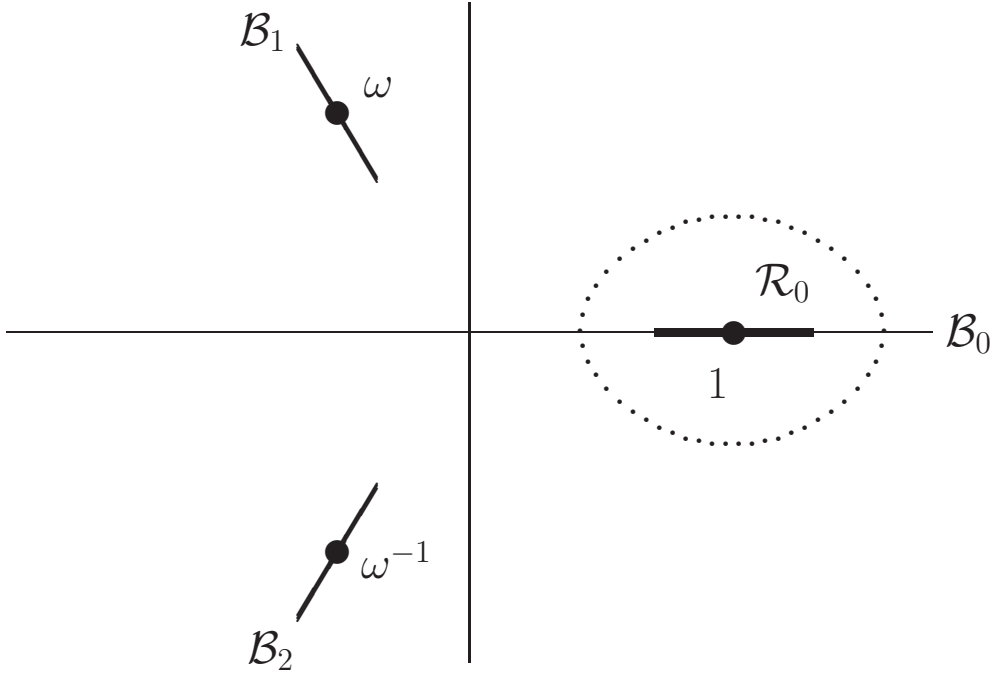
\begin{figure}[hbt]
 \begin{picture}(420,260) (0,0)

 \put (50,125) {\line(1,0) {350}}
 \put (225,0) {\line(0,1) {250}}

 \put (325,125)  {\circle*{9}}
 \put (175,208)  {\circle*{9}}
 \put (175,42)  {\circle*{9}}

 \put (405,120)  {\Large  {${\cal B}_0$}}
 \put (138,234)  {\Large  {${\cal B}_1$}}
 \put (141,0)  {\Large  {${\cal B}_2$}}

 \put (315,100)  {\Large 1}
 \put (185,214)  {\Large $\omega$}
 \put (184,32)  {\Large $\omega^{-1}$}

 \put (333,139) {{\Large {${\cal R}_0$}}}

 \thicklines
 \put (295,124) {\line(1,0) {60}}
 \put (295,125) {\line(1,0) {60}}
 \put (295,126) {\line(1,0) {60}}

 \put (265,125) {\bf .}
 \put (266,131) {\bf .}
 \put (268,137) {\bf .}
 \put (271,142) {\bf .}
 \put (275,147) {\bf .}
 \put (280,152) {\bf .}
 \put (285,156) {\bf .}
 \put (290,159) {\bf .}
 \put (295,162) {\bf .}
 \put (300,164) {\bf .}
 \put (305,166) {\bf .}
 \put (310,167) {\bf .}
 \put (315,168) {\bf .}
 \put (320,168) {\bf .}
 \put (325,168) {\bf .}
 \put (330,168) {\bf .}
 \put (335,167) {\bf .}
 \put (340,166) {\bf .}
 \put (345,164) {\bf .}
 \put (350,162) {\bf .}
 \put (355,159) {\bf .}
 \put (360,156) {\bf .}
 \put (365,152) {\bf .}
 \put (370,147) {\bf .}
 \put (374,142) {\bf .}
 \put (377,137) {\bf .}
 \put (379,131) {\bf .}
 \put (380,125) {\bf .}

 \put (266,119) {\bf .}
 \put (268,113) {\bf .}
 \put (271,108) {\bf .}
 \put (275,103) {\bf .}
 \put (280,98) {\bf .}
 \put (285,94) {\bf .}
 \put (290,91) {\bf .}
 \put (295,88) {\bf .}
 \put (300,86) {\bf .}
 \put (305,84) {\bf .}
 \put (310,83) {\bf .}
 \put (315,82) {\bf .}
 \put (320,82) {\bf .}
 \put (325,82) {\bf .}
 \put (330,82) {\bf .}
 \put (335,83) {\bf .}
 \put (340,84) {\bf .}
 \put (345,86) {\bf .}
 \put (350,88) {\bf .}
 \put (355,91) {\bf .}
 \put (360,94) {\bf .}
 \put (365,98) {\bf .}
 \put (370,103) {\bf .}
 \put (374,108) {\bf .}
 \put (377,113) {\bf .}
 \put (379,119) {\bf .}

 \put (160,16) {\line(3,5) {30}}
 \put (160,17) {\line(3,5) {30}}
 \put (160,18) {\line(3,5) {30}}

 \put (160,234) {\line(3,-5) {30}}
 \put (160,233) {\line(3,-5) {30}}
 \put (160,232) {\line(3,-5) {30}}

 \thinlines

 \ \end{picture}
 \vspace{1.5cm}
 \caption{ The cut $t_p$-plane for $N=3$, showing the three
 branch cuts ${\cal B}_0, {\cal B}_1, {\cal B}_2$ and the 
  approximately circular region ${\cal R}_0$ in which $x_p$ lies 
 when $p \in {\cal D}$.}
 \label{brcuts}
 \end{figure}

 We regard $x_p, y_p, \mu_p^N$ as functions of $t_p$.
 Then $t_p$ lies in a complex plane containing $N$ branch cuts
 ${\cal B}_0, {\cal B}_1, \ldots ,$ $ {\cal B}_{N-1}$ on the lines 
 $\arg (t_p) = 0, 2 \pi /N , \ldots ,$ $ 2 \pi (N-1)/N$,
 as indicated in Fig. \ref{brcuts}, while $x_p$ lies in a 
 near-circular
 region round the point $x_p = 1$, as indicated schematically by the
 region ${\cal R}_0$ inside the dotted curve of Fig. \ref{brcuts}. The 
 variable $y_p$ can lie anywhere in the complex plane {\em except}
 in ${\cal R}_0$ and in $N-1$ corresponding near-circular regions 
 ${\cal R}_1, \ldots ,{\cal R}_{N-1}$ round the  other branch cuts.
 With these choices, we say that $p$ lies in the ``domain'' $\cal D$.

 With these choices, we show in \cite{RJB05a} that
 \be \label{GS}
 G_{pq}(r) \eq k^{(N+1-2r)/N^2} \, S(t_p) \comma \ee
 for $r = 1, \ldots , N-1$, while
 \be
 G_{pq}(0) \eq G_{pq}(N) \eq k^{(1-N)/N^2} \, S(t_p)^{1-N} \period \ee
 Hence $G_{pq}(1) \cdots G_{pq}(N) = 1$, in agreement with the 
 definition (\ref{defG}). The function $S_p = S(t_p)$ is given by
 \be \label{defS}
 \log S(t_p) \eq - \frac{2}{N^2} \log k + \frac{1}{2N\pi} 
 \int_{0}^{2 \pi} \frac{k' \e^{\i \theta} }{1- k' \e^{\i \theta} }
 \, \log[ \Delta (\theta )  - t_p ] \, {\rm d} \theta \comma \ee
 where 
 \be \Delta (\theta ) \eq [(1 - 2 k' \cos \theta +{k'}^2 )/k^2]^{1/N} 
 \period \ee
 {From} \cite{RJB05b}, particular properties are
 \bd S(0) = 1 \sep S(\infty) = k^{-2/N^2} \comma \ed
 \be \label{Sprod}
 S(t_p ) S(\omega t_p) \cdots S(\omega^{N-1} t_p ) \eq 
 k^{-1/N} \, x_p \period \ee
 The function $S(t_p)$ is single-valued, non-zero and analytic in the 
 cut $t_p$ plane of Figure \ref{brcuts}, but only the cut on the 
 positive real axis is necessary: the other cuts can be removed for 
 this function. If $S_{ac}(t_p)$ is the analytic continuation of
 $S(t_p)$ across the branch cut ${\cal B}_r$, then
 \ba \label{Sac}
 S_{ac}(t_p)   \del \del  &  \del   = \; \; S(t_p)   &  
 {\rm for} \; \; r \neq 0 \nonumber \\
 \del & \; \; \; \; \; \; \; \; = \; \; (y_p/x_p) S(t_p)   &  
 {\rm for} \; \; r  = 0 \period \ea

 If we interchange $p,q$ in eqn. 49 of \cite{RJB05a}, then apply 
 the restriction (\ref{qp}) and use the relation $RS = MVRSV$
 together with  eqn. 53 of \cite{RJB05a}, we obtain
 \be G_{pq}(r) G_{p',q'}(N-r) = 1 \comma \ee
 where $p' = V^{-1} q' = RSVp$. It follows that $S(t_p)$ also has the 
 symmetry
 \be \label{SpSp}
 S_p S_{RSVp} = S(t_p) S(1/t_p) \eq k^{-2/N^2} \period \ee
 





 \section{ The Riemann sheets (``domains'') formed by analytic 
  continuation }

 We shall want to consider the analytic continuation of certain 
 functions of $t_p$ onto other Riemann sheets, i.e. beyond the domain 
 $\cal D$. We restrict 
 attention to functions that are meromorphic and single-valued in 
 the cut plane of Figure \ref{brcuts}, and similarly for their 
 analytic continuations. Obvious examples  are $x_p, y_p$ and 
 $S(t_p)$. They are therefore  meromorphic and single-valued 
 on their Riemann surfaces, but we need to know what these 
 surfaces are.

 We start by considering the most general such surface. As a 
 first step,
 allow $\mu_p$ to move from outside the unit circle to inside. Then 
 $t_p$ will cross one of the $N$  branch cuts 
 ${\cal B}_i$ in Figure  \ref{brcuts}, 
 moving onto another Riemann sheet, going back to its original value
 but now with $y_p$ in ${\cal R}_i$.
 Since $y_p$ is thereby confined to  the region near and surrounding
 $\omega^i$, we say that $y_p \simeq \omega^i$. Conversely,
 by $y_p \simeq \omega^i$ we mean that $y_p \in {\cal R}_i$

 We say that $p$ has moved into the  {\em domain}  ${\cal D}_i$
 {\em adjacent} to $\cal D$. There are $N$ such domains
 ${\cal D}_0, {\cal D}_1,$ $\ldots , {\cal D}_{N-1}$.

 Now allow $\mu_p$ to become larger than one, so
 $t_p$ again crosses one of the $N$ branch cuts. Again we require  
 that $t_p$ returns to its original value. If it crosses
 ${\cal B}_i$, then it moves back to the original domain $\cal D$.
 However, if it crosses  another cut ${\cal B}_j$ then 
 $x_p$ moves into ${\cal R}_{j-i}$, 
 and we say that $p$ is now in domain ${\cal D}_{i,j-i}$.

 Proceeding in this way, we build up a Cayley tree of domains.
 For instance, the domain ${\cal D}_{ijk}$ is a third neighbour
 of $\cal D$, linked via the first neighbour ${\cal D}_{i}$ and the
 second-neighbour ${\cal D}_{ij}$, as indicated in Figure
 \ref{seq}. Here $x_p \simeq 1 $ in ${\cal D}$,
 $y_p \simeq \omega^i $ in ${\cal D}_{i}$,
 $x_p \simeq \omega^j $ in ${\cal D}_{ij}$  and 
 $y_p \simeq \omega^k $ in ${\cal D}_{ijk}$. 
 We reject moves that take $p$ back to the domain
 immediately before the last, so $j \neq 0$ and $k \neq i$.
 We refer to the sequence $\{ i,j,k ,\ldots \}$ that define
 any domain as a {\it route}. We can think of it as a sequence of
 points, all with the same value of $t_p$, on the successive
 Reimann sheets or domains.

 The domains $\cal D$, ${\cal D}_{ij}$, ${\cal D}_{ijk\ell}$,...
 with an even number of indices, have $x \simeq \omega^{\ell}$, where
 $\ell$ is the last index. We refer to them as being of even 
 {\it parity} and of {\em type} $\ell$. The domains ${\cal D}_i$, 
 ${\cal D}_{ijk}$,... have $y \simeq \omega^{\ell}$ and are of 
 odd parity and type $\ell$.



 \setlength{\unitlength}{1pt}
 \begin{figure}[hbt]
 \begin{picture}(420,60) (0,30)
 \put (100,25) {\line(1,0) {35}}
 \put (80,20) {\Large ${\cal D}$ }
 \put (150,20) {\Large ${\cal D}_i$}
 \put (175,25) {\line(1,0) {35}}
 \put (220,20) {\Large ${\cal D}_{ij}$}
 \put (250,25) {\line(1,0) {35}}
 \put (295,20) {\Large ${\cal D}_{ijk}$}
 \end{picture}
  \vspace{1.5cm}
 \caption{ A sequence of adjacent domains ${\cal D}, {\cal D}_i, 
 {\cal D}_{ij}, {\cal D}_{ijk}$.}
 \label{seq}
 \end{figure}

 The automorphism that takes a point $p$ in $\cal D$ to 
 a point in  ${\cal D}_i$, respectively, is
 the mapping
 \be
 A_{i} \eq  V^{i-1} \, R V^{-i} \period \ee
 If $q = A_i \, p$, then
 \be \label{amap}
 x_q = \omega^{-i} y_p \sep y_q = \omega^i x_p \sep t_q = t_p 
 \period \ee
 Because of (\ref{Veqns}), $A_{i+N} = A_i$, so there are $N$ such 
 automorphisms.

 We can use these maps to generate all the sheets in the full
 Cayley tree. Suppose we have a domain with route $\{ i,j,k ,\ldots \}$
 and we apply the automorphism $A_{\alpha}$ to all  points on the route. 
 From (\ref{amap}) this will generate a new route
 $\{\alpha, i\minus \alpha,j\plus \alpha,k \minus \alpha ,\ldots \}$.
 For instance, if we apply the map $A_{\alpha}$ to the route  $\{ m \}$
 from ${\cal D}$ to  ${\cal D}_m$, we obtain the route
 $\{ \alpha,  m \minus \alpha \}$ to the domain 
 $D_{ \alpha,  m \minus \alpha }$. Thus the map that takes $\cal D$
 to ${\cal D}_{ij}$ is $A_i A_{i+j}$.

 Iterating, we find that the map that takes $\cal D$
 to ${\cal D}_{ijk \ldots mn}$ is
 \be \label{map1}
 A_i A_{i+j} A_{j+k}  \cdots A_{m+n} \period \ee
 We must have
 \be \label{AB}
 A_i^2 \eq 1 \comma \ee
 since applying the same map twice merely returns $p$ to  
 the previous domain. 

 Let us refer to the general Riemann surface we have just described
 as ${\cal G}$. It consists of infinitely many Riemann sheets, 
 each sheet corresponding to a site on a Cayley tree, 
 adjacent sheets corresponding to adjacent points on the tree.
 A Cayley tree is a huge graph: it contains no circuits and is 
 infinitely dimensional, needing infinitely many integers 
 to specify all its sites.

 Any given function will have a  Riemann surface that can be 
 obtained  from ${\cal G}$ by identifying
 certain sites with one another, thereby creating circuits
 and usually reducing the graph to one of finite dimensionality. 

 {From} (\ref{amap}), the maps $A_0, A_1,\ldots , A_{N-1}$ leave 
 $t_p$ unchanged.
 We shall often find it helpful to regard $t_p$ as a fixed
 complex number, the same in all domains,  and to consider the 
 corresponding values of 
 $x_p, y_p$ (and the hyperelliptic variables $z_p, w_p$) in 
 the  various domains. To within factors of $\omega$, the
 variables  $x_p$ and $y_p$
 will be the same as those for $\cal D$ in even domains, while
 they will be interchanged on odd domains.

 \subsection*{Analytic continuation of $S(t_p)$}

 Now return to considering the function $S(t_p)$. It is sometimes 
 helpful to write this more explicitly as $S(x_p,y_p)$. Then from
 (\ref{Sac}) the map that takes $S(t_p)$ from domain $\cal D$
 to  ${\cal D}_i$ is 
 \be \label{AiS}
 q = A_i p: \; \;   S(x_q,y_q) \eq  (y_q/x_q)^{-\delta_i} 
 \, S(x_p,y_p) \comma \ee
 where $x_q, y_q$ are given by (\ref{amap}) and 
 $\delta_i = 1$ if $i = 0$, mod $N$; otherwise
 $\delta_i = 0$. Note that  $x_q, y_q$ are obtained by interchanging 
 $x_p, y_p$ and multiplying them by powers of $\omega$.

 For given $t_p$, let $S_0(t_p)$ be the value of $S(t_p)$ 
 in the central domain $\cal D$, given by the formula (\ref{defS}).
 Iterating the mappings (\ref{AiS}) from domain to domain, 
 in any domain we must have
 \be \label{Sa}
 S(x_p,y_p) \eq \omega^{\alpha} \, (y_p/x_p)^r\, S_0(t_p) \comma \ee
 where $\alpha$, $r$ are integers. Note that in this equation
 $x_p,y_p$ are the values for the domain being considered: they
 are {\em not} the corresponding initial values of the central domain
 $\cal D$. 

 In particular,
 in the domain ${\cal D}_{ijk}$ we obtain 
 \be \label{Sijk}
 r = -\delta_i + \delta_{i+j} - \delta_{j+k}  \period \ee






 \section{The original hyperelliptic parametrization for $N=3$}

 Hereinafter we restrict our attention to the case $N=3$ and use the
 hyperelliptic parametrization and notation of previous 
 papers.\cite{RJB91,RJB93a,RJB93b,RJB98a}
 We use only formulae that involve ordinary Jacobi
 elliptic (or similar) functions of one variable.

 Given $k, k'$, we 
 define a ``nome'' $x$  by
 \be \label{defx}
 (k'/k)^2 = 27 x \prod_{ n =1}^{\infty} 
 \left( \frac{1-x^{3n}}{1-x^n} \right)^{12} \period \ee
 We regard $x$ as a given constant, {\em not} 
 the same as the rapidity variable $x_p$.
 It is small at low temperatures ($k'$ small), and increases to unity
 at criticality ($k' = 1 $). We introduce two elliptic-type  functions
 \be \label{defh}
 h(z) \eq  \omega^2 \, h(x z) \eq  \prod_{n=1}^{\infty} 
 \frac{ (1-  \omega x^{n-1} z) 
 (1-\omega^2 x^n/z) } { (1-\omega^2 x^{n-1} z) 
 (1- \omega x^n/z) } \comma \ee
 \be \label{defphi}
 \phi(z) \eq z^{1/3}  \prod_{n=1}^{\infty} \frac{(1-x^{3n-2} /z)
 \, (1-x^{3 n-1}z)}{(1-x^{3n-2} z) \, (1-x^{3 n-1}/ z)} 
 \period \ee

 We then define two further variables  $z_p, w_p$ by
 \be \label{eq27}
 t_p = x_p y_p = \omega h(z_p) = h(-1/w_p) 
 = \omega^2 h(- w_p/z_p)  \comma \ee
 These are the relations (27) of \cite{RJB93a}. 
 The relations (32) of \cite{RJB93a} are also satisfied:
 \be  \label{eq32}
 x_p^{-3}  y_p^3 \, \mu_p^{-6} = \phi(x z_p/w_p^2)^3 = 
 \phi(- x z_p w_p)^3  =  \phi(-x w_p/z_p^2)^3 \comma \ee
 as are the relations (4.5), 
 (4.6) of \cite{RJB93b}, in particular


 \be \label{eq4.5}
 w = \prod_{n=1}^{\infty} \frac{(1-x^{2n-1} z/w) (1-x^{2n-1} w/z)
 (1-x^{6n-5} zw) (1-x^{6n-1} z^{-1} w^{-1})} 
 {(1-x^{2n-2} z/w) (1-x^{2n} w/z)
 (1-x^{6n-2} zw) (1-x^{6n-4} z^{-1} w^{-1})} \ee
 writing $z_p, w_p$ here simply as $z, w$.

 The $z_p, w_p$ variables satisfy the automorphisms
 \bd z_{Rp} = x z_p \sep z_{Sp} = 1/(x z_p) \sep z_{Vp} = -1/w_p 
 \comma \ed
 \be\label{zwauto}
 w_{Rp} = z_p/w_p \sep w_{Sp} = 1/(x w_p) \sep w_{Vp} = z_p/w_p 
 \period \ee
 The operation $p \rightarrow M p$ multiplies $(z_p w_p)^{1/3}$ by 
 $\omega$, but does not change $z_p, w_p$ themselves.

 The variables $z_p, w_p$  are of order unity when $k', x$ are small, 
 $\mu_p$ is of order  $1/k'$, and $x_p \simeq 1$. This is the 
 low-temperature limiting case of $p \in {\cal D}$. It is convenient to 
 define 
 \be u_p = \{ z_p, -1/w_p, -w_p/z_p \} \period \ee

 The three automorphisms that leave $t_p$ unchanged, while 
 taking $\cal D$ to 
 ${\cal D}_0, {\cal D}_1, {\cal D}_2$, respectively, are
 \be \label{3autos1}
 A_0 =   V^{2}R \sep A_1 = R V^{2} \sep A_2 = V R V \period \ee
 If $q = A_i \, p$, then
 \be \label{3autos2}
 x_q = \omega^{-i} y_p \sep y_q = \omega^i x_p \sep t_q = t_p 
 \comma \ee
 and $u_q \eq {\cal A}_i u_p$, where ${\cal A}_0, 
 {\cal A}_1, {\cal A}_1$ are the three-by-three matrices
 \bd {\cal A}_0 \eq   \left( \begin{array}{ccc}
 0 & x^{-1} & 0 \\
 x & 0 & 0 \\
 0 & 0 & 1 \end{array} \right)  \sep 
 {\cal A}_1 \eq \left( \begin{array}{ccc}
 0 & 0 & x \\
 0 & 1 & 0 \\
 x^{-1} & 0 & 0 \end{array} \right) 
 \sep 
 {\cal A}_2 \eq \left( \begin{array}{ccc}
 1 & 0 & 0  \\
 0 & 0 & x^{-1} \\
 0 & x & 0 \end{array} \right)
 \ed
 They satisfy the identities 
 \be \label{trip}
 {\cal A}_i  {\cal A}_j  {\cal A}_i \eq  {\cal A}_j  
 {\cal A}_i   {\cal A}_j \ee
 for all $i,j$. 

 They permute the three elements $z_p, -1/w_p, -w_p/z_p$ 
 of $u_p$ and multiply them by powers of $x$, the product of the 
 elements remaining unity.  Let 
 $z_p^0, w_p^0$ be the values of
 $z_p,w_p$ on the central sheet $\cal D$. Then it follows that on 
 any sheet, for the same common value of $t_p$,
 \be \label{intmn}
 z_p = x^m \alpha_p \sep w_p = x^n \beta_p \comma \ee
 where $ \{ \alpha_p, -1/\beta_p, -\beta_p/\alpha_p \}$ is a 
 permutation of  $\{ z_p^0, -1/w_p^0, -w_p^0/z_p^0 \}$.

 Repeated applications of the three automorphisms 
 will therefore generate a two-dimensional set of permutations and 
 multiplications of the elements of $u_p$. Each member of the set 
 corresponds to a site on the honeycomb lattice
 of Figure \ref{honeylatt}. Adjacent Riemann sheets correspond
 to adjacent sites of the lattice. Sheets of even parity
 correspond to sites represented by circles, those of odd 
 parity are represented by squares.
 if $i$ is the integer inside the circle or square, then
 for even sites $x_p \simeq \omega^{-i}$, while on odd sites
 $y_p \simeq \omega^{i}$. The numbers shown in brackets 
 alongside each site are the integers $m,n$ of (\ref{intmn}).

 Thus for the functions $z_p$ and $w_p$ of $t_p$, the graph 
 $\cal G$ of the Riemann surface reduces to this 
 two-dimensional honeycomb lattice.

 Note that the sites $X, Y, Z$ in the figure are third neighbours
 of the central site $\cal D$, and each can be reached from
 $\cal D$ in two three-step ways. For instance, $Y$ is both
 ${\cal D}_{021}$ and ${\cal D}_{211}$.\footnote{Note that 
 for ${\cal D}_{ijk}$ we here take the intermediate site $j$ to be 
 represented Figure \ref{honeylatt} by the integer $-j$, mod 3.
 This is changed in the next section to $+j$.}

 Thus $Y$ is obtainable from $\cal D$ by the maps $A_0 A_2 A_0$ and
 $A_2 A_0 A_2$. From (\ref{trip}) these are the same, so we can
 identify the two sheets as one, represented by $Y$.

 Similarly, $X$ corresponds to $A_1 A_2 A_1 = A_2 A_1 A_2$
 and   $Z$ to $A_0 A_1 A_0 = A_1 A_0 A_1$. This is why
 $\cal G$ reduces to the honeycomb lattice.

 {\em However}, the automorphisms (\ref{AiS})
 of the function $S(t_p)$ do {\em not} in general satisfy (\ref{trip}). 
 On its Riemann  sheets ${\cal D}_{021}$, ${\cal D}_{211}$
 we find from (\ref{Sijk})  that $r = - 2$ and $1$.
 Thus from (\ref{Sa}) the analytic continuation of $S(t_p)$
 is $ (y_p/x_p)^{-2} S_0(t_p)$ and $ (y_p/x_p) S_0(t_p)$ on each sheet, 
 respectively (ignoring factors of $\omega$). Thus 
 the result for $Y$ depends on the route taken to it.
 (The same is true of $Z$, but not for $X$.)

 Hence the values of $z_p$ (and $w_p$) are the same on 
 ${\cal D}_{021}$ and ${\cal D}_{211}$, but $S(t_p)$ is different.
 It follows that  $z_p$, $w_p$ do not uniquely determine  
 $S(t_p)$. Hence neither $S(t_p)$ nor  $S(t_p)^3$ is a single-valued 
 function  of these hyperelliptic variables $z_p, w_p$: one must look 
 elsewhere for such a parametrization.


 \setlength{\unitlength}{1pt}
 \begin{figure}[hbt]

 \begin{picture}(420,260) (0,0)

 \put (196,105) {\circle{15}}
 \put (193,101) {0}
 \put (203,105)  {\line(1,0) {44}}
 \put (190,111)  {\line(-3,5) {22}}
 \put (190,99)  {\line(-3,-5) {22}}
 \put (163,101) {\protect \footnotesize (0,0) }
 \put (196,83) {\large ${\cal D}$}

 \put (107,155) {\circle{15}}
 \put (104,151) {1}
 \put (114,155)  {\line(1,0) {44}}
 \put (101,161)  {\line(-3,5) {22}}
 \put (101,149)  {\line(-3,-5) {22}}
 \put (74,151) {\protect \footnotesize (1,2) }

 \put (107,55) {\circle{15}}
 \put (104,51) {2}
 \put (114,55)  {\line(1,0) {44}}
 \put (101,61)  {\line(-3,5) {22}}
 \put (101,49)  {\line(-3,-5) {22}}
 \put (74,51) {\protect \footnotesize (2,1) }

 \put (287,155) {\circle{15}}
 \put (284,152) {1}
 \put (294,155)  {\line(1,0) {44}}
 \put (281,161)  {\line(-3,5) {22}}
 \put (281,149)  {\line(-3,-5) {22}}
 \put (248,152) {\protect \footnotesize (-2,-1) }

 \put (287,55) {\circle{15}}
 \put (284,52) {2}
 \put (294,55)  {\line(1,0) {44}}
 \put (281,61)  {\line(-3,5) {22}}
 \put (281,49)  {\line(-3,-5) {22}}
 \put (248,52) {\protect \footnotesize (-1,-2) }

 \put (196,5) {\circle{15}}
 \put (193,2) {1}
 \put (203,5)  {\line(1,0) {44}}
 \put (190,11)  {\line(-3,5) {22}}
 \put (160,2) {\protect \footnotesize (1,-1) }

 \put (196,205) {\circle{15}}
 \put (193,202) {2}
 \put (203,205)  {\line(1,0) {44}}
 \put (190,199)  {\line(-3,-5) {22}}
 \put (160,202) {\protect \footnotesize (-1,1) }

 \put (158,148) {\line(1,0) {15}}
 \put (158,162) {\line(1,0) {15}}
 \put (158,148) {\line(0,1) {14}}
 \put (173,148) {\line(0,1) {14}}
 \put (163,151) {2}
 \put (178,151) {\protect \footnotesize (0,1) }

 \put (158,48) {\line(1,0) {15}}
 \put (158,62) {\line(1,0) {15}}
 \put (158,48) {\line(0,1) {14}}
 \put (173,48) {\line(0,1) {14}}
 \put (163,51) {1}
 \put (178,51) {\protect \footnotesize (1,0) }

 \put (247,-2) {\line(1,0) {15}}
 \put (247,12) {\line(1,0) {15}}
 \put (247,-2) {\line(0,1) {14}}
 \put (262,-2) {\line(0,1) {14}}
 \put (252,1) {2}
 \put (267,1) {\protect \footnotesize (0,-2) }
 \put (244,18) {\large $Z$}

 \put (247,98) {\line(1,0) {15}}
 \put (247,112) {\line(1,0) {15}}
 \put (247,98) {\line(0,1) {14}}
 \put (262,98) {\line(0,1) {14}}
 \put (252,101) {0}
 \put (267,101) {\protect \footnotesize (-1,-1) }

 \put (247,198) {\line(1,0) {15}}
 \put (247,212) {\line(1,0) {15}}
 \put (247,198) {\line(0,1) {14}}
 \put (262,198) {\line(0,1) {14}}
 \put (252,201) {1}
 \put (267,201) {\protect \footnotesize (-2,0) }
 \put (244,182) {\large $Y$}

 \put (338,148) {\line(1,0) {15}}
 \put (338,162) {\line(1,0) {15}}
 \put (338,148) {\line(0,1) {14}}
 \put (353,148) {\line(0,1) {14}}
 \put (343,151) {2}
 \put (358,151) {\protect \footnotesize (-3,-2) }

 \put (338,48) {\line(1,0) {15}}
 \put (338,62) {\line(1,0) {15}}
 \put (338,48) {\line(0,1) {14}}
 \put (353,48) {\line(0,1) {14}}
 \put (343,51) {1}
 \put (358,51) {\protect \footnotesize (-2,-3) }

 \put (66,98) {\line(1,0) {15}}
 \put (66,112) {\line(1,0) {15}}
 \put (66,98) {\line(0,1) {14}}
 \put (81,98) {\line(0,1) {14}}
 \put (71,101) {0}
 \put (86,101) {\protect \footnotesize (2,2) }
 \put (49,101) {\large $X$}

 \put (66,198) {\line(1,0) {15}}
 \put (66,212) {\line(1,0) {15}}
 \put (66,198) {\line(0,1) {14}}
 \put (81,198) {\line(0,1) {14}}
 \put (71,201) {1}
 \put (86,201) {\protect \footnotesize (1,3) }

 \put (66,-2) {\line(1,0) {15}}
 \put (66,12) {\line(1,0) {15}}
 \put (66,-2) {\line(0,1) {14}}
 \put (81,-2) {\line(0,1) {14}}
 \put (71,1) {2}
 \put (86,1) {\protect \footnotesize (3,1) }

 \end{picture}

 \vspace{1.5cm}
 \caption{\footnotesize The honeycomb lattice formed by the 
 hyperelliptic variables $z, w$ in the alternative parametrization.
 Circles (squares) denote sites of even (odd) parity.}
 \label{honeylatt}
 \end{figure}





 \section{The alternative hyperelliptic parametrization for $N=3$}

 There is another way of parametrizing $k, x_p, y_p, \mu_p, t_p$
 so that the nome $x$ is small when $k'$ is small. It can be obtained
 from the original  parametrization of the previous section by a simple 
 mapping, and we do this in the Appendix. We take the results
 (\ref{defxa}), (\ref{eq27a}), (\ref{eq32a}), (\ref{zwautosa})
 therein and drop the hats on $x_p, y_p, \ldots, V, M$ to obtain

 \be \label{defx5}
 -{k'}^2 = 27 x \prod_{n =1}^{\infty} 
 \left( \frac{1-x^{3n}}{1-x^n} \right)^{12} \period \ee
 \be \label{eq275}
 y_p/x_p = \omega h(z_p) = h(-1/w_p) 
 = \omega^2 h(- w_p/z_p)  \comma \ee
 \be  \label{eq325}
 - x_p^{-3}  y_p^3 \, \mu_p^{-6} = 
 \phi(x z_p/w_p^2)^3 = 
 \phi(- x z_p w_p)^3   =  \phi(-x w_p/z_p^2)^3 \period \ee

 \bd z_{Rp} = -x w_p \sep z_{Sp} = -1/(x w_p) \sep
 z_{Vp} = -1/ w_p \comma \ed
 \be \label{zwautos5}
 w_{Rp} = w_p/z_p \sep w_{Sp} = -1/(x z_p) \sep
 w_{Vp} = z_p/ w_p \period \ee
 Again $z_{Mp} = z_p$, $w_{Mp} = w_p$. These equations 
 replace (\ref{defx}), (\ref{eq27}), (\ref{eq32}), 
 (\ref{zwauto}) of section 4.
 The functions $h(z)$, $\phi(z)$ remain defined by
 (\ref{defh}) and (\ref{defphi}), and the relation
 (\ref{eq4.5}) remains satisfied.

 We now regard  $x_p, y_p, \ldots, V, M$ as being the same 
 variables and  automorphisms as those above,
 satisfying  (\ref{prlns}), (\ref{autos}) and (\ref{Veqns}). 
 Then the
 hyperelliptic variables $x, z_p, w_p$ are {\em different} from 
 those of section 4. If one takes $p \in {\cal D}$ and expands 
 the functions 
 in powers of the low-temperature variable $k'$,
 then to leading order $z_p, w_p$ are the same  
 as the $z_p, w_p$ of section 4, being of order unity and satisfying
 $w_p = z_p+1$; $x$ is negated and is of order $k'$.

 The three automorphisms that leave $t_p$ unchanged while taking 
 $\cal D$ to ${\cal D}_0$,
 ${\cal D}_1$, ${\cal D}_2$ are again given by (\ref{3autos1})
 and  (\ref{3autos2}). Using the rules (\ref{zwautos5}),
 we find that $z_p, w_p$ transform according the rules
 \be \label{Aalt}
 q = A_i p \, : \; \; \;  
 z_q = x^{2-i-3\delta (i)}/z_p \sep w_q = x^{i-1}/w_p \comma \ee
 for $i = 0, 1, 2$, writing $\delta_i$ as $\delta (i)$.
 If $z_p^0, w_p^0$ are the values of $z_p, w_p$ on the central 
 sheet $\cal D$, then it follows that on any Riemann sheet 
 the analytic continuations of $z_p, w_p$ (for a given value of $t_p$) 
 are
 \be \label{anzw}
 x^m \, \left( z_p^0 \right)^{\pm 1} \sep x^n \, 
 \left( w_p^0 \right)^{\pm 1} \comma \ee
 choosing the upper (lower) signs on sheets of even (odd) parity.
 Here  $m,n$ are integers satisfying
 \ba 
 m+n & = &  0 \; \; \; ( {\rm mod} \; 3 ) \; \; {\rm on \; \;
 even \; \; sheets } \comma \nonumber \\
 & = &  1 \;  \; \; ({\rm mod} \; 3 ) \; \; {\rm on \; \;
 odd \; \; sheets } \period \ea
 The Riemann surface
 for $z_p, w_p$ therefore corresponds to a two-dimensional
 graph $\cal G$, each site of  $\cal G$ being specified by the 
 two integers $m, n$.

 In fact this $\cal G$ is the honeycomb lattice shown in Figure
 \ref{honeylatt}, but we must interpret it slightly differently
 from what we did in section 4. Adjacent sites still correspond to 
 adjacent Riemann sheets, and the parities of the sites are  
 shown as in section 4 by circles and squares. Again, on odd sites
 $y_p \simeq \omega^i$, where $i$ is the number shown in the Figure.
 However, on even sites we now take $x_p \simeq \omega^i$.
 For each site, the bracketed integers shown in Figure \ref{honeylatt} 
 are now the integers $(m,n)$ of (\ref{anzw}).

 As in the previous section, the sites $X, Y, Z$ in the figure 
 are third neighbours
 of the central site $\cal D$, and each can be reached from
 $\cal D$ in two three-step ways. There is the difference that
 the $j$ of ${\cal D}_{ijk}$ is now the number inside the
 corresponding circle of Figure \ref{honeylatt}, so in this section
 $X$ is both ${\cal D}_{210}$ and ${\cal D}_{120}$, so from (\ref{map1})
 and (\ref{AB})
 it is obtainable from $\cal D$ by the maps $A_2 A_0 A_1$ and
 $A_1 A_0 A_2$.  The reason we can identify
 ${\cal D}_{210}$ with ${\cal D}_{120}$ is that
 (\ref{Aalt}) implies that $A_2 A_0 A_1 = A_1 A_0 A_2$.
 More generally, it implies that
 \be \label{Aijk}
 A_i A_j A_k \eq A_k A_j A_i \ee
 for all permutations $i,j,k$ of $0,1,2$. This means that
 $X, Y, Z$ each corresponds to a single Riemann sheet rather than two, 
 and is the reason that $\cal G$ reduces (for the functions $z_p, w_p$
 of $t_p$) from the full Cayley tree to the honeycomb lattice.

 \subsection*{Properties of $z_p, w_p, S(t_p)$}

 Within the central domain $\cal D$ there is some circle of 
 non-zero radius, centre the origin, such that none of $z_p, w_p,$
 $1/z_p, 1/w_p,$ $z_p/w_p, w_p/z_p$ can lie within the circle
 (for $x$ small the radius is of order $x^{1/2}$). Two special 
 values of $p$ that lie within  $\cal D$ are
 \bd 
 p(1): \; \; \; z_p = -1/w_p = -w_p/z_p = \omega^2  \sep y_p =  0 
 \ed 
 \be \label{p1}
 x_p = k^{1/3} \sep S(t_p) = 1 \comma  \ee
 and 
 \bd 
 p(2): \; \; \; z_p = -1/w_p = -w_p/z_p = \omega   \sep y_p = \infty 
 \ed 
 \be \label{p2} x_p = k^{-1/3} \sep S(t_p) = k^{-2/9} \comma \ee
 using eqns. (\ref{prlns}) and (\ref{Sprod}).

 Remembering that $x_p \simeq 1$, one can verify that the 
 conditions (\ref{eq4.5}), (\ref{eq275}) are satisfied. These are 
 the only two points within $\cal D$ where $y_p/x_p$ has a 
 zero or pole.

 Any point where one of $z_p, -1/w_p, -w_p/z_p$ 
 is equal to $\omega^2 x^m$ or $\omega x^m$, for non-zero integer $m$, 
 necessarily lies {\em outside} $\cal D$ (i.e. on another Riemann 
 sheet). It follows that the functions $G(z_p), \ldots,  h(-w_p/z_p)$
 defined and used in the next section have no zeros or poles
 for $p \in \cal D$: they  are finite and non-zero therein.
 The same is true of $x_p$, $S(t_p)$ and the function 
 $\tilde{S}(z_p,w_p)$ defined below.


 Now consider the maps $A_i$ for the function $S(t_p)$, as given in
 (\ref{AiS}) - (\ref{Sijk}). The mapping  $A_i A_j A_k$ simply 
 multiplies $S(t_p)$ by a power of $\omega$ and by $(y_p/x_p)^r$, 
 where $r = -\delta_i + \delta_j - \delta_k$. Obviously this 
 exponent $r$ is unchanged by interchanging $i$ with $k$, so 
 is the same for both sides of (\ref{Aijk}).

 The powers of $\omega$ are {\em not} necessarily the same, but
 we can avoid this difficulty by simply working with 
 $S(t_p)^3$ instead of $S(t_p)$. The automorphisms $A_i$ for 
 $S(t_p)^3$ do satisfy (\ref{Aijk}) and $S(t_p)^3$
 is the same on the two sheets  ${\cal D}_{210}$, ${\cal D}_{120}$, 
 so again $X$ reduces to a single sheet. Similarly, so do 
 $Y$ and $Z$. 

 As a result,  $S(t_p)^3$ is {\em uniquely determined} if we 
 know $z_p$ and $w_p$. In fact on any Riemann sheet $(m,n)$ the analytic 
 continuation of $S(t_p)^3$ is
 \be \label{ancS}
 (y_p/x_p)^{3r} S_0(t_p)^3 \comma \ee
 where
 \ba \label{rmn}
 r & = &  (m+n)/3 \; \; {\rm on \; \; even \; \; sheets } \comma 
 \nonumber \\
 & = &  (m+n-1)/3 \; \; {\rm on \; \; odd \; \; sheets } \period \ea
 The $x_p, y_p$ in (\ref{ancS}) are those of the sheet 
 under consideration.

 It therefore makes sense to look for a single-valued
 meromorphic function of $z_p$ and $w_p$ that is equal to $S(t_p)^3$. 
 We can write $S(t_p)$ itself as $S(z_p,w_p)$ provided we accept
 that it is three-valued in the rather trivial way that its values 
 differ by factors of $\omega$. Then from the automorphisms 
 (\ref{AiS}) and (\ref{Aalt}), this function must satisfy the three 
 relations
 \be  \label{autosSzw}
 S(z_p, w_p ) = (x_p/y_p) S(x^{-1} z_p^{-1}, x^{-1} w_p^{-1} )
 = S(x \, z_p^{-1}, w_p^{-1} ) = S(z_p^{-1}, x \, w_p^{-1} )  \ee
 for all $z_p, w_p$. Also, using (\ref{autos}) and (\ref{zwautos5}),
 the relation (\ref{Sprod}) becomes
 \be \label{prrelnS}
 S(z_p,w_p) S(-1/w_p, z_p/w_p) S(-w_p/z_p, -1/z_p) \eq k^{-1/3} x_p 
 \period \ee

 The function $x_p^{-1}$ has no zeros or poles on even sheets, whereas on 
 odd sheets (where $y_p \simeq \omega^i$) it has the 
 same zeros and poles as $y_p/x_p$. it is therefore useful to work not 
 with $S(z_p, w_p )^3$, but the function
 \be \label{deftS}
 \tilde{S} (z_p, w_p ) \eq x_p^{-1}  S(z_p, w_p )^3 \comma \ee
 since from   (\ref{ancS})  and (\ref{rmn}) this has the same poles 
 and zeros as $(y_p/x_p)^{m+n}$ on {\em all} Riemann sheets $(m,n)$, 
 even and  odd. From (\ref{eq275}) we can write $(y_p/x_p)$
 as a function of either $z_p, w_p$ or $z_p/w_p$: this suggests that
 it may be possible to write $\tilde{S} (z_p, w_p )$ as a product
 of functions of these individual variables. We do this in the next 
 section.





 \section{$S(t_p)$ as a function of $z_p, w_p$}

 Define the functions
 \bd F(z)  \eq \prod_{j=1}^{\infty} \frac{(1-x^j z)^j}
 {(1-x^j z^{-1}  )^j}  \ed
 \be \label{defG3}
 G(z) = F(\omega z) /F(\omega^2 z) \comma \ee
 so $F(z) \eq 1/F(z^{-1})$, $G(z) \eq G(z^{-1})$ and
 \be \label{Greln}
 G(z)/G(x z)\eq \omega h(z)  \ee
 for all complex number $z$. 

 Consider 
 the product 
 \be \label{defP}
 P  \eq G(z_p)^{\alpha} \, G(-1/w_p)^{\beta} \, G(-w_p/z_p)^{\gamma} \ee
 for arbitrary integers $\alpha, \beta, \gamma$.
 As a first step, we ask if we can choose $\alpha, \beta, \gamma$ so 
 that $P$ has the same poles and zeros
 as $\tilde{S} (z_p, w_p )$, i.e. as $(y_p/x_p)^{m+n}$  on all sheets
 $(m,n)$.

 The integers $m,n$ specify a sheet and are defined in (\ref{anzw}).
 {From} (\ref{eq275}), on sheet $(m,n)$ the function $y_p/x_p$ has a 
 simple zero when
 $z_p = x^m \omega^2$, $w_p = - x^n \omega$, 
 $z_p/w_p = -\omega^{m-n} \omega$. Any one of these equalities 
 implies the other two, so at this point $\tilde{S} (z_p, w_p )$
 has a zero of order $m + n$. On the other hand,
 the three factors of $P$ have zeros of order $-m\alpha, n\beta, 
 (m-n)\gamma$, respectively. Thus we require
 $m+n = -m\alpha+n\beta+(m-n)\gamma$, for all allowed integers 
 $m,n$ and fixed values of $\alpha, \beta, \gamma$. This will be so 
 iff
 \be \label{Pconds} \alpha = \gamma - 1 \sep \beta = \gamma + 1  
 \period \ee

 There are also possible zeros and poles when
 $z_p = x^m \omega$, $w_p = - x^n \omega^2$, 
 $z_p/w_p = -\omega^{m-n} \omega^2$, but the only difference
 from the above is that all the orders are negated, so again we
 obtain the conditions (\ref{Pconds}).

 Neither $\tilde{S} (z_p, w_p )$ not $P$ has any other other zeros or 
 poles, and  $\tilde{S} (z_p, w_p )$ is independent of the integer 
 $\gamma$, which is still arbitrary. Substituting (\ref{Pconds})
 into (\ref{defP}), this implies that the functions
 \bd
 \tilde{S} (z_p, w_p )  G(z_p)/ G(-1/w_p) \sep
 G(z_p) G(-1/w_p) G(-w_p/z_p)   \ed
 have no zeros or poles anywhere on the Riemann surface. Using 
 (\ref{p1}) and (\ref{deftS}),  they have values
 $k^{-1/3},  F(\omega)^{-3}$ at $p = p(1)$, so 
 if they were constants it would follow that
 \be \label{ident1}
 \tilde{S} (z_p, w_p ) \eq k^{-1/3}G(-1/w_p) / G(z_p) \comma \ee
 \be  \label{ident2}
 G(z_p) G(-1/w_p) G(-w_p/z_p)  \eq 1/F(\omega)^{3} \period \ee

 We can prove that these relations are indeed true by using Liouville's
 theorem for a single Riemann sheet.  Let $T(z_p,w_p)$ 
 be the ratio of 
 the RHS of (\ref{deftS}) to the RHS of (\ref{ident1}). Then from
 (\ref{autosSzw}) and (\ref{defG3}) - (\ref{Greln}) it follows that
 \be 
 T(z_p, w_p ) = T(x^{-1} z_p^{-1}, x^{-1} w_p^{-1} )
 = T(x \, z_p^{-1}, w_p^{-1} ) = T(z_p^{-1}, x \, w_p^{-1} )  
 \period \ee
 Thus the function $T$ is unchanged by the three automorphisms
 $A_0, A_1, A_2$. It is therefore a single-valued function of
 the variable $t_p$, {\em without} the branch cuts of Figure 
 \ref{brcuts}. It has no zeros or poles in ${\cal D}$, so 
 it has no zeros or poles in the complex $t_p$ plane, including the 
 point
 at infinity. By Liouville's theorem it is therefore a constant.
 It is unity when $z_p = \omega^2$ or $\omega$, i.e. when 
 $t_p = 0$ or $\infty$, so it is one. This proves the identity
 (\ref{ident1}). The identity (\ref{ident2}) can be proved 
 similarly by taking  $T(z_p,w_p)$ to  be the ratio of of
 the LHS of (\ref{ident2}) to the RHS.
 
 \subsection*{Two more identities}

 We originally tried a much more general ansatz for the product $P$, 
 allowing for factors such as $1-x^j z$ raised to a power linear in
 $j$ and  $\mod(j,3)$. As a result we discovered yet two more 
 identities satisfied by $z_p, w_p$.
 Define  the function
 \be \tilde{h}(z) \eq \prod_{n=1}^{\infty} \frac{(1-x^{3n-2} \omega z ) 
 (1-x^{3n-1} \omega^2 /z)}
 {(1-x^{3n-2} \omega^2 z ) (1-x^{3n-1} \omega / z)} \comma \ee
 then we find that
 \be \label{ident3}
 \tilde{h}(z_p) \, \tilde{h}(-w_p) \eq  \tilde{h}(-1/w_p)\,  
 \tilde{h}(-z_p/w_p) \eq  \tilde{h}(-w_p/z_p)
 \,  \tilde{h}(1/z_p)  \period \ee

 We can prove these identities in a similar way. First note that
 \bd \tilde{h}(x/z) = 1/\tilde{h}(z) \sep  \tilde{h}(z/x) = 
 \omega^2/\tilde{h}(x^{-1} z^{-1}) \ed 
 and
 \bd h( z) \eq  \tilde{h}(z/x) \,   \tilde{h}(z) \, \tilde{h}(1/z) 
 \ed
 for all $z$. We can use this last formula and (\ref{eq275}) to 
 eliminate ratios such as 
 $\tilde{h}(-x^{-1} w^{-1})/\tilde{h}(x^{-1} z)$ in favour of
 $\tilde{h}$ functions whose arguments do not contain $x^{-1}$ as a 
 factor. Using this 
 fact and applying the automorphisms $A_0, A_1, A_2$ to the 
 ratios of the expressions in (\ref{ident3}),
 we find that the automorphisms merely permute these ratios.
 If we write the three expressions as $J_1,J_2,J_3$ and form,
 for arbitrary $\alpha$,
 \bd T(z_p, w_p) \eq (\alpha - J_2/J_1)(\alpha - J_3/J_2)
 (\alpha - J_1/J_3)(\alpha - J_1/J_2)(\alpha - J_2/J_3)
 (\alpha - J_3/J_1) \comma \ed
 then this $T(z_p, w_p)$ is unchanged by the automorphisms, so 
 is a single-valued function of $t_p$. It has no zeros or poles
 in $\cal D$, so by Liouville's theorem it is a constant.
 At $p = p(1)$ or $p(2)$, the $J_i$'s are equal, so
 for all $z_p, w_p$
 \bd  T(z_p, w_p) \eq (\alpha-1)^6 \period \ed
 It follows that $J_1 = J_2 = J_3$ for all $t_p$, which 
 establishes the identities (\ref{ident3}).

 For an arbitrarly chosen numerical value of $z_p$, working to 
 32 digits of accuracy, we have successfully checked 
 the identities (\ref{ident1}),  (\ref{ident2}),
 (\ref{ident3}) to twenty terms in an expansion in powers of $x$.


 {From} (\ref{autos}) and (\ref{zwautos5}), the map 
 $p \rightarrow RSVp$ takes $x_p,t_p,z_p,w_p$ to 
 $1/x_p$, $1/t_p$, $-w_p$, $-z_p$.
 {From} (\ref{deftS}) and (\ref{ident1}), noting that 
 $G(-1/w_p) = G(-w_p)$, it is apparent that
 \bd S_p S_{RSVp} \eq S(z_p,w_p) S(-w_p,-z_p) \eq k^{-2/9} \comma \ed
 in agreement with (\ref{SpSp}).




 \section{Summary}

 For the case when $N=3$ and $q$ is related to $p$ by (\ref{qp}), the 
 generalised order parameter function $G_{pq}(r)$ is given by
 (\ref{GS}), where $x_p^{-1} S(t_p)^3$ can be simply expressed by  
 (\ref{deftS}) and (\ref{ident1})  as a product of functions of the
 hyperelliptic variables $z_p, w_p$ of section 5. This is 
 the first time that a thermodynamic property of the $N>2$ chiral 
 Potts model has been so expressed. (As distinct from algebraic 
 functions of the Boltzmann  weights, such as the function of 
 $f_{pq}$ of \cite{RJB98a}.)

 The functions $F(z)$, $G(z)$ in (\ref{defG3}) are infinite products
 similar to elliptic functions, except that factors such as $1-x^j z$
 are  raised to the power $j$. Such extensions of elliptic functions
 occur in the free energies of other solvable models, notably
 the Ising model.  Good examples are the equations (B.10), (B.17) and 
 (B.18) of \cite{RJB03}.

 
 We emphasize that  (\ref{deftS}) and (\ref{ident1}) are in 
 terms of the {\em alternative} hyperelliptic parametrization of 
 section 5 herein. Papers \cite{RJB91} to \cite{RJB05c} are in
 terms of the original hyperelliptic parametrization of 
 section 4.

 It is still an interesting question whether  $G_{pq}(r)$
 can be simply expressed as a function of such variables for 
 arbitrary $p,q$. The result of this paper implies that 
 one must use the hyperelliptic parametrization
 of section 5, rather than that of section 4. There is a difficulty with 
 this: if we write $R^6 p, R^6 q$ as $p', q'$, the relations 
 (\ref{zwautos5}) imply that
 $z_{p'} = z_p$ and $w_{p'} = w_p$, whereas it is {\em not} true
 that   $G_{p', q}(r) =  G_{pq}(r)$ or that
 $G_{p, q'}(r) =  G_{pq}(r)$. This means that $G_{pq}(r)$
 cannot be a single-valued function of $z_p, w_p, z_q, w_q$.
 However, the function $L_{pq}(r) = G_{pq}(r) G_{Rq,Rp}(r)$
 of \cite{RJB98} is unchanged by $p \rightarrow p'$, and 
 by $q \rightarrow q'$, so may be so expressible.





 \section*{Appendix}

 \setcounter{equation}{0}
 \renewcommand{\theequation}{A\arabic{equation}}

 Here we show how the alternative hyperelliptic parametrization 
 of section 5  can be obtained from the original  
 parametrization of section 4 by a simple mapping.

 Let $k, k', x_p, y_p,\mu_p,  t_p$ be the variables of section 4.
 Define new variables $\hat{k}, {\hat{k}}', \hat{x}_p, \hat{y}_p, 
 \hat{\mu}_p, \hat{t}_p$ so that
 \bd k = \hat{k}^{-1}  \sep k' = \i \hat{k}'/\hat{k} \sep 
 x_p =1/\hat{x}_p  \sep y_p = \hat{y}_p  \comma \ed
 \be \mu_p  = \e^{-\i \pi/2 N} \hat{x}_p \hat{\mu}_p \sep 
 y_p/ x_p = \hat{t}_p   \period  \ee
 Leave $x, z_p, w_p$ and the functions $h(z)$, $\phi(z)$
 unchanged.

 Then the relations (\ref{kk'}), (\ref{prlns}) remain satisfied if we 
 replace $k, k', x_p, y_p, \mu_p, t_p$ therein by 
 $\hat{k}, {\hat{k}}', \hat{x}_p, \hat{y}_p, 
 \hat{\mu}_p, \hat{t}_p$. The relations (\ref{defx}), 
 (\ref{eq27}),  (\ref{eq32}) become
 \be \label{defxa}
 -{\hat{k}}'^2 = 27 x \prod_{N=1}^{\infty} 
 \left( \frac{1-x^{3n}}{1-x^n} \right)^{12} \period \ee
 \be \label{eq27a}
 \hat{y}_p/ \hat{x}_p = \omega h(z_p)  = h(- 1 /w_p)
 = \omega^2 h(- w_p/z_p,x)
 \comma \ee
 \be  \label{eq32a}
 - \hat{x}_p^{-3}  \hat{y}_p^3 \, \hat{\mu}_p^{-6} = 
 \phi(x z_p/w_p^2)^3 = 
 \phi(- x z_p w_p)^3  =  \phi(-x w_p/z_p^2)^3 \period \ee

 Define automorphisms $\hat{R}, \hat{S}, \hat{V}, \hat{M}$
 by  (\ref{autos}) with $x_p, y_p,  \ldots , M_p$ replaced by
 $\hat{x}_p, \hat{y}_p, \ldots , \hat{M}_p$.  Then
 \bd \hat{R} = V S \sep \hat{S} = M^{-1} V^{-1} R 
 \sep \hat{V} = V \sep \hat{M} = M  \ed
 and the relations (\ref{Veqns}) remain satisfied if 
 $R, S, V, M$ therein are replaced by 
 $\hat{R}, \hat{S}, \hat{V}, \hat{M}$. 
 {From} (\ref{zwauto}) it follows that
 \bd z_{\hat{R}p} = -x w_p \sep z_{\hat{S}p} = -1/(x w_p) \sep
 z_{\hat{V}p} = -1/ w_p \comma \ed
 \be \label{zwautosa}
 w_{\hat{R}p} = w_p/z_p \sep w_{\hat{S}p} = -1/(x z_p) \sep
 w_{\hat{V}p} = z_p/ w_p \period \ee

 Now drop the hats on $\hat{k}, {\hat{k}}', \hat{x}_p, \ldots , 
 \hat{M} $ to obtain the equations (\ref{defx5}) - (\ref{zwautos5}) of
 section 5. Equations (\ref{kk'}), (\ref{prlns}), (\ref{autos}) 
 remain true.


 \end{document}